\documentclass[%
 reprint,
 amsmath,amssymb,
 aps,
]{revtex4-2}

\usepackage{graphicx}
\usepackage{dcolumn}
\usepackage{bm}
\usepackage{hyperref}
\hypersetup{colorlinks=true,
linkcolor=cyan,
urlcolor=blue,
citecolor=red}
\usepackage{ulem}
\usepackage{comment}
\usepackage{amssymb}
\usepackage{tabularx} 

\makeatletter
\newcommand{\Rmnum}[1]{\expandafter\@slowromancap\romannumeral #1@} 
\makeatother

\begin{document}
\preprint{APS/123-QED}

\title{Beyond Kasner Epochs: Ordered Oscillations and Spike Dynamics Inside Black Holes with Higher‑Derivative Corrections}

\author{\textbf{Mei-Ning Duan}$^{1,2}$}
\author{\textbf{Li Li}$^{1,2,3}$}
\author{\textbf{Yu-Xuan Li}$^{1,2}$}%
\author{\textbf{Fu-Guo Yang}$^{2,4}$}%

\vspace{1cm}

\affiliation{${}^{1}$Institute of Theoretical Physics, Chinese Academy of Sciences, Beijing 100190, China}
\affiliation{${}^{2}$School of Physical Sciences, University of Chinese Academy of Sciences,
Beijing 100049, China}
\affiliation{${}^{3}$School of Fundamental Physics and Mathematical Sciences, Hangzhou Institute for Advanced Study, UCAS, Hangzhou 310024, China}
\affiliation{${}^{4}$International Centre for Theoretical Physics Asia-Pacific,
University of Chinese Academy of Sciences, 100190 Beijing, China}

\begin{abstract}
{Building upon the long-standing paradigm that dynamics near a spacelike singularity are governed by a sequence of Kasner epochs, we demonstrate that this picture is fundamentally altered when higher-curvature or quantum gravitational corrections are included. By incorporating such terms alongside a minimally coupled scalar field, we discover three distinct dynamical phases near the singularity: modified Kasner eons, persistent periodic oscillations, and oscillatory spike dynamics with growing amplitude. In particular, the Kasner-like geometry persisting only in highly constrained situations. The latter two regimes represent a clean departure from classical Kasner phenomenology, revealing a richer and more ordered landscape of behaviors in the deep interior of black holes beyond Einstein gravity. This work establishes a comprehensive approach for understanding the gravitational nonlinearity in the most extreme gravitational environment.}

\end{abstract}

\maketitle

\textbf{Introduction.--}Spacetime singularities inside a black holes are a generic feature in classical general relativity (GR) of gravity. At the singularity, spacetime curvature becomes infinite, and all known laws of physics break down. This is not just an incompleteness of the theory but a signpost pointing toward a deeper theory—quantum gravity. Understanding the nature of spacetime singularities is central to resolving foundational problems in physics, including the preservation of causality and the black hole information paradox. Notably, the initial Big Bang singularity in classical cosmology underscores that probing black hole singularities also illuminates the origins of our universe. The dynamics near spacelike singularities in GR are famously described by the Belinski-Khalatnikov-Lifshitz (BKL) limit~\cite{Lifshitz:1963ps,Belinsky:1970ew,Belinskii:1973sud,Belinsky:1981vdw}, which predicts a chaotic, oscillatory approach dominated by successive Kasner epochs. For large classes of theories, the evolution of Kasner exponents can be understood as a particle moving on a hyperbolic billiard table, known as Cosmological Billiards~\cite{Damour:2002et}. More recently, a plethora of rich near-singularity dynamics have been uncovered by coupling matter fields to the original gravitational theory, for which a common feature is the emergence of Kasner scaling and bounces between different Kasner epochs towards the singularity, see \emph{e.g.}~\cite{Hartnoll:2022rdv,Cai:2023igv,Cai:2024ltu,DeClerck:2023fax}. 

However, as the singularity is approached, quantum corrections will become increasingly important. Although there is still lack of a complete quantum gravity theory, one ubiquitous manifestation of quantum effect is higher-derivative correction to the classical equations of motion~\cite{tHooft:1974toh,Gross:1986mw,Sakharov:1967nyk}. These corrections trigger new dynamical phases called Kasner eons~\cite{Bueno:2024fzg}: a Kasner-like geometry dominated by emergent physics at each energy scale. Given that matter are ubiquitous and could exert a substantial backreaction on the geometry, a comprehensive understanding of the Kasner eon physics necessitates a systematic investigation of the combined interplay between higher-derivative gravitational corrections and dynamic matter fields. Recently, the case with a free scalar with quadratic potential was consider in~\cite{Caceres:2024edr}, modifying the Kasner exponents compared to the vacuum case~\cite{Bueno:2024fzg,Bueno:2024qhh}.
Nevertheless, what lies beyond the quadratic potential is completely unknown, raising a fundamental question: what physical principles, if any, govern the interior structure of black holes?

The celebrated BKL conjecture and its “Cosmological Billiards” description provide a powerful framework for the chaotic approach to spacelike singularities in classical GR. However, a systematic understanding of singularity dynamics in the presence of higher-derivative corrections—a ubiquitous ingredient in quantum-gravitational effective theories—remains absent. This gap is crucial, as such terms become dominant in the extreme UV regime near the singularity. Previous studies of “Kasner eons” have focused on vacuum or free-scalar cases, leaving the interplay between generic scalar potentials and higher-curvature terms unexplored. Our work fills this gap by establishing a comprehensive framework. We demonstrate that the classical BKL paradigm is dramatically supplanted by three novel dynamical phases. This reveals that quantum-gravitational corrections can in fact impose order, leading to periodic or structured oscillatory behavior deep inside black holes.


\textbf{Setup and Kasner eons.}--We consider a minimally coupled scalar field $\psi$ on top of the gravity theory with a tower of higher-curvature corrections in $(d+1)$-spacetime dimensions:
\begin{equation}\label{HighDerivGrav}
    \mathcal{L}=R-\frac{1}{2}(\partial\psi)^2-V(\psi)+\sum_{N\geq 2}\alpha_N\mathcal{L}_{R_{(N)}}\,,
\end{equation}
where the first term (Ricci scalar $R$)
constitutes the standard Einstein-Hilbert action and encodes the dynamics of classical GR. $\mathcal{L}_{R_{(N)}}$ are purely constructed out of curvature invariants of $N$-th curvature order, while $\alpha_N$ are coupling coefficients with dimensions of length$^{2(N-1)}$. The generalized Einstein equation reads
\begin{equation}\label{eomg}
{{G}^{\mu}}_{\nu}+\sum_{N=2}^{N_{\text{max}}}\alpha_N{{G_{(N)}}^{\mu}}_{\nu}={{T}^{\mu}}_{\nu}\,,
\end{equation}
with ${{G}^{\mu}}_{\nu}$ the Einstein tensor and ${{G_{(N)}}^{\mu}}_{\nu}$ the contribution from $\mathcal{L}_{R_{(N)}}$. 

To illustrate the significant challenges posed by higher-curvature terms, we reconsider the Cosmological Billiards picture for the case with Gauss-Bonnet (GB) term $\mathcal{L}_{R_{(2)}}=R_{\mu\nu\alpha\beta}R^{\mu\nu\alpha\beta}-4R_{\alpha\beta}R^{\alpha\beta}+R^2$, which is ubiquitous in string-theoretic completions of gravity\,\footnote{We employ the notation for the Riemann curvature tensor, ${R^{\alpha}}_{\mu\beta\nu}=\partial_\beta {\Gamma^\alpha}_{\nu\mu}-\partial_\nu {\Gamma^\alpha}_{\beta\mu}+{\Gamma^\alpha}_{\beta\sigma}{\Gamma^\sigma}_{\nu\mu}-{\Gamma^\alpha}_{\nu\sigma}{\Gamma^\sigma}_{\beta\mu}$, with the Ricci tensor defined as its contraction $R_{\mu\nu}={R^{\rho}}_{\mu\rho\nu}$.}. Under the affine parameter diagonal matrix ansatz~\cite{Damour:2002et},
\begin{equation}\label{liyxdiagansatz}
    ds^2=-e^{-2\beta(t)}dt^2+\sum_{i=1}^d e^{-2\beta^i(t)}dx_i^2,\quad \beta=\sum_{i=1}^{d}\beta^i\,,
\end{equation}
one has the effective Lagrangian
\begin{equation}\label{cosbill}
    \begin{split}
    \mathcal{L}&= (\delta_{ij}-1){{\frac{d\beta}{dt}}^{i}}{{\frac{d\beta}{dt}}^{j}}+\frac{1}{2}({\frac{d\psi}{dt}})^2-e^{-2\beta}V(\psi)\\
    &\qquad\qquad\quad+\frac{\alpha_2}{3} e^{2\beta}T_{i j k l}{{\frac{d\beta}{dt}}^{i}}{{\frac{d\beta}{dt}}^{j}}{{\frac{d\beta}{dt}}^{k}}{{\frac{d\beta}{dt}}^{l}}\,,
    \end{split}
\end{equation}
with $T_{ijkl}=(\delta_{ij}-1)(\delta_{i k}-1)(\delta_{i l}-1)(\delta_{j k}-1)(\delta_{j l}-1)(\delta_{k l}-1)$. The first line represents the free Kasner (billiard) motion in superspace $(\beta^i, \psi)$ valid when the scalar potential $V$ is negligible. This dynamics is radically altered by the velocity-dependent terms arising from higher-curvature corrections.
Interestingly, the term in the second line of~\eqref{cosbill} is reminiscent of Lagrangians for classical time crystals~\cite{Shapere:2012nq}. In the following, we show how this term, combined with matter couplings, introduces key novel physical effects.

Given the inherent complexities introduced by higher-derivative terms, we begin our investigation by examining the hairy black holes within a plane-symmetric ansatz.
\begin{equation}
\label{AdSMetric}
\mathrm{d}s^2=\frac{1}{z^2}\left[-f(z)\mathrm{e}^{-\chi(z)}\mathrm{d}t^2+\frac{\mathrm{d}z^2}{f(z)}+\mathrm{d}\vec{x}_{d-1}^2\right],\;\psi=\psi(z)\,,
\end{equation}
where the singularity is located at $z\rightarrow\infty$ and the event horizon is $z_H$ at which $f(z_H)=0$. 
Note that for specific coupling values (\emph{e.g.} negative GB coefficient), higher-curvature terms can induce a finite-volume curvature singularity~\cite{Bueno:2024fzg}, a case we reserve for future study~\footnote{Note, however, that the GB term must have a nonnegative coefficient by considering a tree-level ultraviolet completion free of ghosts or tachyons~\cite{Cheung:2016wjt}.}. The Kasner eons resulting from~\eqref{AdSMetric} can be written in terms of proper time $\tau$:
\begin{equation}\label{KasnerMetric}
\begin{split}
\mathrm{d}s^2&=-\mathrm{d}\tau^2+C_t\tau^{2p_t}\mathrm{d}t^2+C_x\tau^{2p_x}\mathrm{d}\vec{x}_{d-1}^2\,,\\
\end{split}
\end{equation}
with $C_t$ and $C_x$ constants. The geometry~\eqref{KasnerMetric} is fully characterized by two Kasner exponents $p_t$ and $p_x$. As approaching the singularity $\tau\rightarrow 0$, a curvature invariant of order $N$ grows as $R_{(N)}\sim 1/\tau^{2N}$ on metric~\eqref{KasnerMetric}. This scaling forces an inevitable dominance of higher-derivative terms in the extreme UV regime. 

The energy-momentum tensor ${{T}^{\mu}}_{\nu}$ associated with~\eqref{KasnerMetric} reads
\begin{equation}\label{Tmunu}
\begin{split}
{{T}^{t}}_{t}&={{T}^{x}}_{x}=\frac{1}{4}(-2V(\psi)+(\frac{\mathrm{d}\psi}{\mathrm{d}\tau})^2)\,,\\
{{T}^{\tau}}_{\tau}&=\frac{1}{4}(-2V(\psi)-(\frac{\mathrm{d}\psi}{\mathrm{d}\tau})^2)\,.
\end{split}
\end{equation}
Combining with~\eqref{eomg}, the necessary conditions for the metric to be expressible in Kasner form are as follows:
\begin{equation}\label{KasCon}
{G^{t}}_{t}+\sum_{N=2}^{N_{\text{max}}}\alpha_N{{G_{(N)}}^{t}}_{t}={G^{x}}_{x}+\sum_{N=2}^{N_{\text{max}}}\alpha_N{{G_{(N)}}^{x}}_{x}\,.
\end{equation}
Then, we reduce the full equations of motions (EoMs) to equations governing the exponents $p_t$ and $p_x$ at different orders. The existence of Kasner eons is determined by the curvature invariants with the highest curvature order ($\propto\tau^{-2N_{\text{max}}}$). However, it is also necessary to check the subdominant part of~\eqref{KasCon} to have a stable Kasner eon.

For the scalar field to influence the Kasner eon, ${{T}^{\mu}}_{\nu}$ must match the order of $R_{(N)}$, from which one generally has
\begin{equation}\label{T}
(\frac{\mathrm{d}\psi}{\mathrm{d}\tau})^2=2({{T}^{x}}_{x}-{{T}^{\tau}}_{\tau})\sim \frac{1}{\tau^{2N}}\,.
\end{equation}
Therefore, the scalar $\psi$ to a  higher-order eon with $N\geq 2$ exhibits a leading-order asymptotic behavior $\psi \propto \tau^{-N+1}$. However, when $\mathcal{O}(V(\psi))>\mathcal{O}(\psi^{\frac{2N}{N-1}})\sim1/\tau^{2N}$, the energy-momentum tensor~\eqref{Tmunu} diverges even faster than the geometry on the left side of~\eqref{eomg}. Therefore, the dynamical evolution is drastically altered, resulting in the nonlinear patterns beyond the Kasner eons. Note that the quadratic potential $V(\psi)\sim\psi^{2}$ is consistently subdominant for Kasner eons~\eqref{KasnerMetric}, aligning with the findings of~\cite{Caceres:2024edr}. Nevertheless, for a potential $V(\psi)$ with a term diverging faster than $\psi^2$, no Kasner eon can be sustained as the order of the curvature invariant increases. For more details, please refer to the Supplementary Material~\footnote{The Supplementary Material presents the technical details for solving the EoMs and analyzing interior dynamics, thereby supporting the results made in the main text.}.

\textbf{Near-singularity dynamics in Einstein-GB gravity.}--We demonstrate the above analysis with a specific model—the Einstein-GB gravity. The generalization to other cases, such as Lovelock gravities, is straightforward (see the Supplementary Material). One has $N=2$ for the GB case. Depending on whether $V(\psi)$ diverges faster than $\psi^4$, we obtain three distinct dynamics of near-singularity limits. To visualize the different Kasner eons as approaching the singularity, we introduce the effective Kasner exponents:
\begin{equation}\label{EffectiveP}
\widetilde{p}_t=\frac{f}{zf'}(z\chi'-\frac{zf'}{f}+2),\quad \widetilde{p}_x=\frac{2f}{zf'}\,,
\end{equation}
where the prime denotes the derivative with respect to the coordinate $z$. When the geometry~\eqref{AdSMetric} matches a proper Kasner eon near the singularity, $\widetilde{p}_t$ and $\widetilde{p}_x$ reduce to the Kasner exponents $p_t$ and $p_x$ of the metric~\eqref{KasnerMetric} precisely.

\textbf{(i) Kasner eons with subdominant potential.}
This is the case for which the  contribution from $V(\psi)$ is subleading compared to its kinetic term. Then, one has ${T^{t}}_{t} = {-T^{\tau}}_{\tau} = {T^{x}}_{x}$ at leading order from~\eqref{Tmunu}. Matching the dominant terms in the EoMs yields two Kasner-like solutions.
\begin{itemize}
    \item Vacuum case: $p_x=4/d$, $p_t=4/d-1$. The scalar field $\psi$ is a constant and thus we recover the Kasner eon with pure gravity~\cite{Bueno:2024fzg,Bueno:2024qhh}.
    \item Non-vacuum case: $p_t=p_x=2/d$. The scalar field diverges asymptotically as $\psi \sim 1/\tau$. Further, to maintain consistency in the power counting, the potential must diverge slower than $\psi^4$. This is consistent with recent work of~\cite{Caceres:2024edr} where the free massive scalar was studied (see also~\cite{Grandi:2021ajl}).
\end{itemize}

\begin{figure}[hpt]
    \centering
    \includegraphics[width=0.455\textwidth]{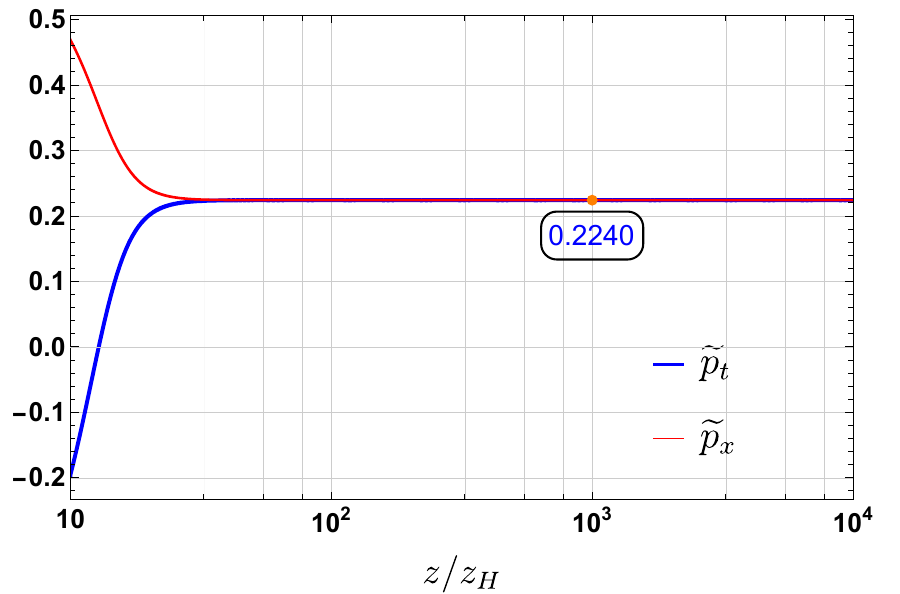}
    \includegraphics[width=0.47\textwidth]{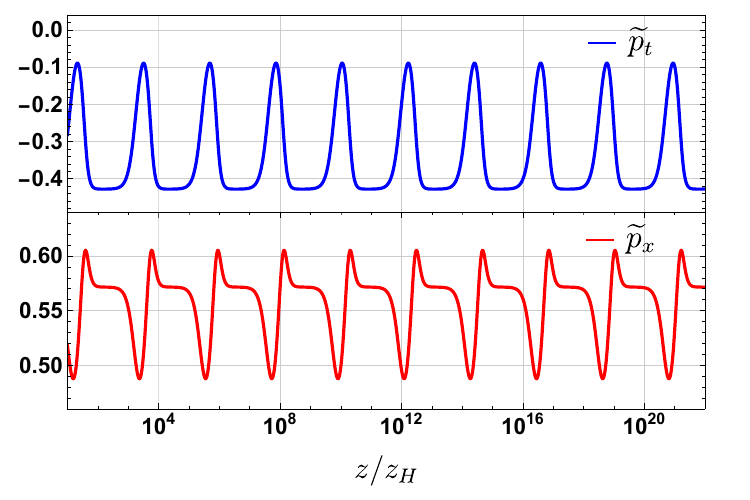}
    \caption{Effective Kasner exponents for hairy Einstein-GB black hole interiors in $d=7$ and $\alpha_2=10^{-2}$ with $\psi(z_H)=0.10$. \textbf{Upper panel}: $V(\psi)=-42-3\psi^2-2\psi ^4$, for which the plateau corresponds to a hairy GB eon of Case I with $p_t=p_x=0.224$. \textbf{Bottom panel}: $V(\psi)=-42-3\psi^2+2\psi ^4$, where the Kasner eon is replaced by periodic oscillations.}
    \label{fig:case12} 
\end{figure}

\textbf{(ii) Critical scenario with quartic potential.}
The critical case is obtained when the potential is quartic in $\psi$, its contribution becomes comparable to the kinetic term, for which ${T^{t}}_{t}={T^{x}}_{x}\neq{-T^{\tau}}_{\tau}$. By matching dominant contributions to the EoMs, we find two solutions for Kasner eon of~\eqref{KasnerMetric} with $\psi\sim 1/\tau$. To check if the Kasner eon is stable or not, we need to consider the subdominant part of~\eqref{KasCon}, which reads ${G^{t}}_{t}={G^{x}}_{x}$ for the GB case.
\begin{itemize}
    \item Case I: $p_t=p_x$. The behavior is similar to the non-vacuum case with subdominant potential. We still have $\psi\sim 1/\tau$, but the value of $p_t$ ($p_x$) is no longer uniquely fixed by the spacetime dimension (see~\eqref{eqpx} below). Moreover, to match the orders of the equation, the potential must satisfy $\mathcal{O}(V(\psi))\sim\mathcal{O}(\psi^4)$. One can show that ${G^{t}}_{t}={G^{x}}_{x}$ in this case, resulting in a stable hairy GB Kasner eon. The formation of this modified Kasner eon is demonstrated in the top panel of Fig.~\ref{fig:case12}.
    
    \item Case II: $p_t=3-(d-1)p_x$. While Case II preserves the scalar divergence $\psi\sim 1/\tau$ and requires the balance $\mathcal{O}(V(\psi))\sim\mathcal{O}(\psi^4)$, the Kasner eon geometry is destabilized. The breakdown is driven by the divergent anisotropy $|{{G}^{t}}_{t}-{{G}^{x}}_{x}|=|2(dp_x-3)|/\tau^2$. Consequently, although a transient Kasner eon may appear initially, it is inevitably disrupted as approaching the singularity ($\tau\rightarrow 0$). The deviation of $p_x$ from $3/d$ controls the timescale of this breakdown: a larger deviation leads to an earlier collapse of the Kasner phase. To understand what's the new geometry, we have to return to the full EoMs with $V\sim c_4 \psi^4$. Interestingly, it turns out to be a completely new type of periodic oscillatory behavior,  illustrated in the bottom panel of Fig.~\ref{fig:case12}. 
\end{itemize}

Based on our numerical results for the quartic potential $V\sim c_4\psi^4$, which show $f\gg1$ and $\psi\gg1$ near the singularity in the critical case, the late‑time equations can be recast in terms of $\rho=\ln z$. Neglecting subleading terms yields (see the Supplementary Material)
\begin{equation}\label{cEoM1}
\begin{split}
&-2C_{\alpha_2} (\frac{\dot{f}}{f}-\frac{d}{2}-\frac{\dot{\chi}}{2})=c_4 (\frac{\psi^2}{f})^2\,,\\
&-2C_{\alpha_2}\dot{\chi}(\frac{\psi}{\dot{\psi}})^2=\frac{\psi^2}{f}\,,\\
&\frac{\ddot{\psi}}{\psi}+(\frac{\dot{f}}{f}-d-\frac{\dot{\chi}}{2})\frac{\dot{\psi}}{\psi}=4 c_4 \frac{\psi^2}{f}\,,
\end{split}
\end{equation}
with $C_{\alpha_2}=(d-1)(d-2)(d-3)\alpha_2$ and the dot denoting $\rho$-derivative.
Despite their highly non‑linear nature, the equations~\eqref{cEoM1} exhibit a key simplification: the left‑hand sides depend only on the logarithmic derivatives $\dot{\psi}/{\psi}$, $\dot{f}/f$, and $\dot{\chi}$, while the right‑hand sides share a common factor $\psi^2/f$. For other forms of $V$, such consistency is absent. Hence, if $\psi^2/f$ is periodic with period $\rho_0$, the solutions can also be periodic,
\begin{equation}
(\frac{\dot{f}}{f}\,, \frac{\dot{\psi}}{\psi}\,, \dot{\chi})|_\rho=(\frac{\dot{f}}{f}\,, \frac{\dot{\psi}}{\psi}\,, \dot{\chi})|_{\rho+\rho_0} \,,
\end{equation}
and the effective Kasner exponents for the new coordinate,
\begin{equation}\label{EffectiveP2}
\begin{split}
\widetilde{p}_t=\frac{f}{\dot{f}}(\dot{\chi}-\frac{\dot{f}}{f}+2),\quad \widetilde{p}_x=\frac{2f}{\dot{f}}\,,
\end{split}
\end{equation}
are manifestly periodic under $\rho\rightarrow\rho+\rho_0$.

One can readily verify that in Case I, the functions $\dot{\psi}/\psi$, $\dot{f}/f$, $\dot{\chi}$ and $\frac{\psi^2}{f}$ are all constant, represents a trivial case where the period $\rho_0$ is arbitrary, \emph{i.e.} the hairy Kasner eon in the top panel of Fig.~\ref{fig:case12}. In contrast, Case II is non-trivial and is characterized by a finite period $\rho_0$
that depends explicitly on the system parameters, as evidenced by the periodic oscillations in the bottom panel of Fig.~\ref{fig:case12}. 
\begin{figure}[hpt]
    \centering
    \includegraphics[width=0.47\textwidth]{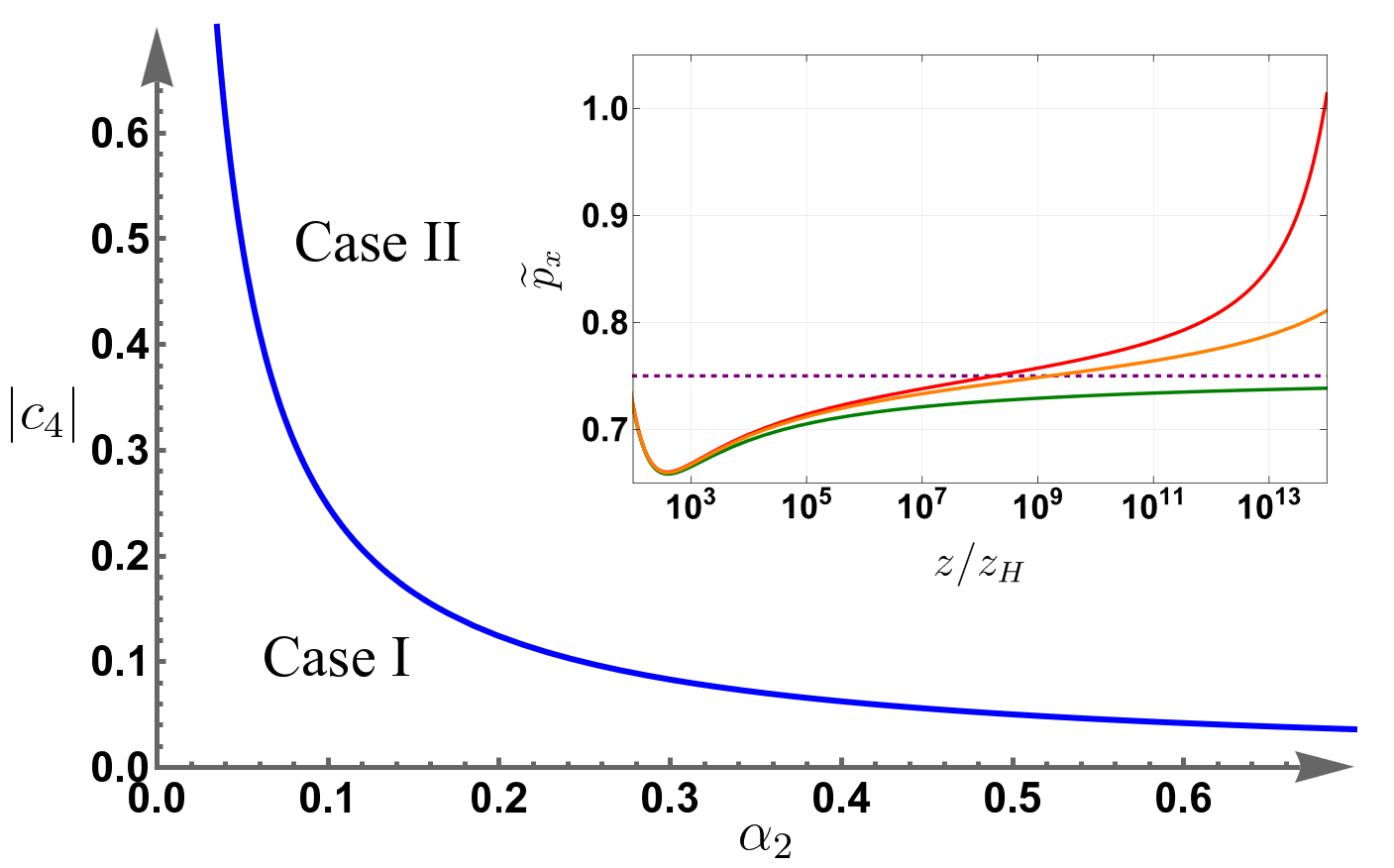}
    \caption{Phase diagram in the plane of GB coupling $\alpha_2$ versus $c_4$ for quartic potential $V\sim c_4 \psi^4$ and $d=4$. The interior dynamics separate into modified Kasner eons (Case I) and periodic oscillations (Case II), with the boundary given by~\eqref{pbc4}. The Case I region shrinks for larger $d$. The inset displays the evolution of $\widetilde{p}_x$  for $V=-12-\frac{3}{2}\psi^2+c_4\psi^4$ with $\psi(z_H)=0.30$ and $\alpha_2=10^{-2}$, where curves from bottom to top corresponds to $c_4=2.47$, 2.485 and 2.49. The purple dashed line marks the critical boundary at $p_x=3/4$, near which a transient Kasner-like regime emerges from Case II side.}
    \label{fig:Trans} 
\end{figure}

The central task is to determine the boundary between Case I and Case II. Although Case II does not admit a stable Kasner eon, a transient Kasner-like regime can emerge initially if the effective exponent $\widetilde{p}_x$ close to $3/d$. This is demonstrated in the inset of Fig.~\ref{fig:Trans} for $d=4$: the green curve develops a plateau with $\widetilde{p}_x\approx 0.732$ around $z/z_H\sim10^{10} $, signaling a temporary Kasner eon. As $z/z_H$ increases, deviations grow and eventually destroy this structure. Nevertheless, the state where $\widetilde{p}_x\approx 3/d$ provides a well-defined reference point for fixing the boundary, as the Kasner geometry~\eqref{KasnerMetric} remains applicable there.

For Case I where $p_t=p_x$, the complete EoMs yield
\begin{equation}\label{eqpx}
c_4\alpha_2=-\frac{2-d p_x}{16{p_x}^3(d-1)(d-2)(d-3)}\,,
\end{equation}
which has a maximum value at
\begin{equation}\label{pbc4}
 c_4^*\alpha_2^*=\frac{d^3}{432(d-1)(d-2)(d-3)}\,,  
\end{equation}
when $p_x=3/d$.
It suggests that for Case I the final Kasner eon dynamics are independent of initial values, but completely control by $c_4\alpha_2$. 
Meanwhile, for Case II with $p_t=3-(d-1)p_x$, one finds, near $p_x=3/d$, that
\begin{equation}\label{CaseII:alpha}
c_4\alpha_2=\frac{1}{16{p_x}^3(d-1)(d-2)(d-3)(4-d p_x)}\,.
\end{equation}
Interestingly, it attains the minimum $c_4^*\alpha_2^*$ precisely at $p_x=3/d$. Thus, one has a sharp phase boundary~\eqref{pbc4} in the $c_4$-$\alpha_2$ plane (Fig.~\ref{fig:Trans}), separating Case I dynamics ($c_4\alpha_2< c_4^*\alpha_2^*$) from Case II ($c_4\alpha_2> c_4^*\alpha_2^*$). Consequently, as the phase boundary is approached from the Case II side, the oscillation period $\rho_0$ increases and ultimately diverges at the boundary.

\begin{figure}[hpt]
    \centering
    \includegraphics[width=0.47\textwidth]{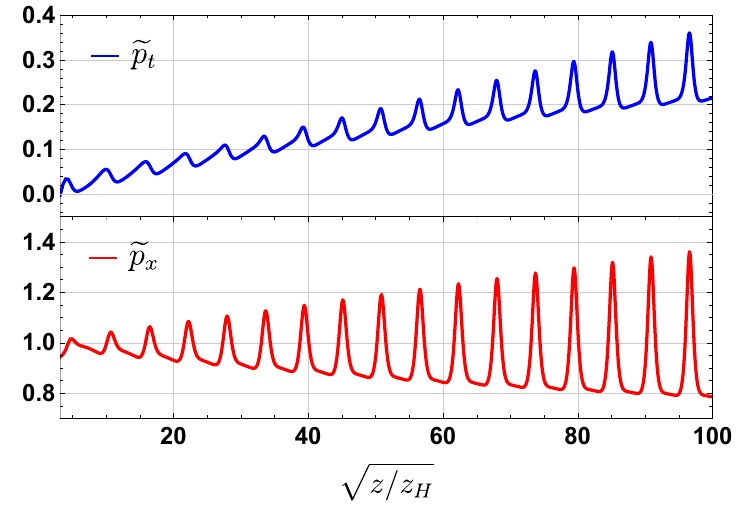}
    \includegraphics[width=0.47\textwidth]{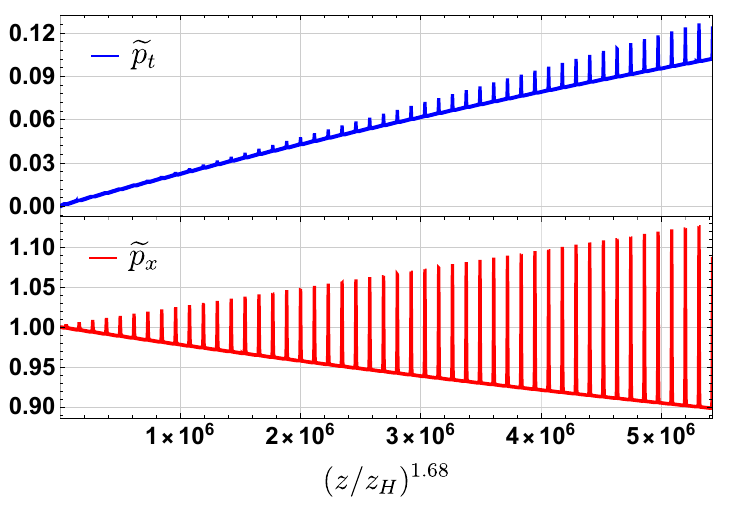}
    \caption{Far‑interior evolution of $\widetilde{p}_t$ and $\widetilde{p}_x$ for $d=4$ and $\alpha_2=10^{-2}$. In contrast to a Kasner plateau, they display oscillatory growth. \textbf{Upper panel}: Potential $V(\psi)=-12-\frac{3}{2}\psi^2+2\psi ^6$ with $\psi(z_H)=0.3$. The peak positions share a common period in the scaled coordinate $({z/z_H})^{0.50}$, while their amplitudes follow a well‑defined envelope. \textbf{Bottom panel}: Potential $V(\psi)=-12-\frac{3}{2}\psi^2+\frac{1}{10}\cosh(\psi)$ with $\psi(z_H)=0.001$. Here the peaks are periodic in $({z/z_H})^{1.68}$.}
    \label{fig:n=6_caseII} 
\end{figure}

\textbf{(iii) Spike dynamics driven by scalar potential.} For potential functions that diverge more rapidly than $\psi^4$, one anticipates even richer dynamics. Indeed, it is clear to see that for $\mathcal{O}(V(\psi))>\mathcal{O}(\psi^4)$, Kasner eon~\eqref{KasnerMetric} can not be a consistent solution as the scalar potential diverges faster than the kinetic term and the GB term. Thus, the interior dynamics should be beyond a Kasner eon and will be sensitive to the details of $V(\psi)$. A potential unbounded from below is widely recognized as a source of serious theoretical pathologies—a ``theoretical catastrophe”. We find that such a potential unavoidably leads to a singularity reached at finite volume. Crucially, this is a feature already inherent to Einstein gravity, rather than a consequence of the higher-derivative corrections (see the Supplementary Material). Thus, we focus on the scalar potential that has a lower bound.

The typical interior dynamics for $V\sim \psi^6$ are shown in the top panel of Fig.~\ref{fig:n=6_caseII}. In contrast to a Kasner eon or periodic oscillation, both effective exponents $\widetilde{p}_t$ and $\widetilde{p}_x$ now exhibit sustained oscillatory growth as the singularity is approached, revealing a possible discrete self-similarity. Notably, their peak oscillations share a common period in the coordinate $\sqrt{z/z_H}$, though their equilibrium positions and amplitudes differ, signaling coupled quasi-periodic evolution. We further note that the precise form of this spike dynamics is sensitive to both the horizon value $\psi(z_H)$ and the detailed shape of $V(\psi)$, yielding an extremely rich phenomenology. The bottom panel of Fig.~\ref{fig:n=6_caseII} illustrates the case with $V\sim \cosh(\psi)$, for which the peaks are periodic in $({z/z_H})^{1.68}$. Additional data on periods and envelopes across models are provided in the Supplementary Material. This observed complexity, in sharp contrast to a Kasner eon, motivates a deeper analytic understanding. Finally, we highlight that the oscillatory behavior found here is distinct from the spike solution discussed in~\cite{Garfinkle:2020lhb}, which arises in geometries with lower symmetry.

\textbf{Conclusion.--}We have developed a unified framework for the near-singularity dynamics of black holes in higher-derivative gravities coupled to a scalar field. Our analysis uncovers a rich phase structure that transcends the classical BKL description: (i) modified Kasner eons, (ii) persistent periodic oscillations, and (iii) oscillatory spike dynamics.  While our concrete analysis focuses on Einstein-GB gravity, the key features generalize to theories with higher powers of curvature, including Lovelock~\cite{Lovelock:1971yv,Maeda:2011ii} and quasi-topological gravities~\cite{Caceres:2024edr,Bueno:2024eig}. The governing criterion is the asymptotic scaling of the scalar potential relative to $\psi^{\frac{2N_{\text{max}}}{N_{\text{max}}-1}}$ where $N_{\text{max}}$ is the highest curvature order. A Kasner eon is possible only when $\mathcal{O}(V(\psi))\leq\mathcal{O}(\psi^{\frac{2N_{\text{max}}}{N_{\text{max}}-1}})$. When $V(\psi)$ diverges faster than this threshold, spike dynamics emerges, with the potential playing the dominant role. Under the critical scenario $V\sim \psi^{\frac{2N_{\text{max}}}{N_{\text{max}}-1}}$ and with suitable couplings, the system exhibits stable periodic oscillations that are reminiscent of classical time crystals. Further examples are presented in the Supplementary Material.

There are various issues worthy of further investigation. First, our findings overturn the conventional understanding of the dynamics near singularities. From the perspective of effective field theory, after accounting for quantum corrections, higher-order curvature terms become dominant at progressively higher energy scales as the singularity is approached.  Consequently, for potentials $V(\psi)$ containing terms that diverge faster than $\psi^2$, any Kasner eon is rendered invalid near the singularity. Instead, a general spike dynamics emerges, which poses a significant challenge to the BKL framework and begs an analytic understanding. Second, in the toy model presented here, we have considered the interior of a static and planar symmetric black hole, which is blind the ultralocal nature of spacetime. A detailed study of less symmetric near-singularity solutions would yield crucial insights into Mixmaster chaos. Moreover, we have illustrated the rich dynamics by the minimally coupled scalar, it will be interesting to consider other matter sector.  Finally, our findings would apply to examining cosmological singularities, where the observed periodic dynamics may suggest novel early-universe scenarios~\cite{Feng:2018qnx,Khodabakhshi:2023sba}.\\

\textbf{Acknowledgments.}--This work is supported by the National Natural Science Foundation of China Grants No.\,12525503, No. 12588101, and No.\,12447101. We acknowledge the use of the High Performance Cluster at Institute of Theoretical Physics, Chinese Academy of Sciences. We acknowledge the use of the High Performance Cluster at the Institute of Theoretical Physics, Chinese Academy of Sciences.

\bibliography{reference}%




\section*{Supplementary material}
\renewcommand{\theequation}{S\arabic{equation}}
\setcounter{equation}{0} 

\renewcommand{\thefigure}{S\arabic{figure}}
\setcounter{figure}{0}


In this Supplementary Material, we provide technical details supporting the results reported in the main Letter. We show the interior dynamics for scalarized Einstein-GB black holes in Section~\ref{SM-GB}. The generalization to higher-order Lovelock gravities can be found in Section~\ref{SM-LL}. In Section~\ref{SM-FS}, we discuss the develop of  finite-volume singularity driven by scalar potentials that lack a lower bound. 

\section{Interior dynamics for scalarized Einstein-GB black holes}\label{SM-GB}
We present more details for the black hole interior dynamics in the Einstein-GB gravity minimally coupled to a real scalar field. The Lagrangian reads
\begin{equation}\label{app:GB_lag}
    \begin{split}
    \mathcal{L}&=R-\frac{1}{2}(\partial_\mu\psi)^2-V(\psi)+\mathcal{L}_{GB}\,,\\
     \mathcal{L}_{GB}&=\alpha_2 (R_{\mu\nu\alpha\beta}R^{\mu\nu\alpha\beta}-4R_{\alpha\beta}R^{\alpha\beta}+R^2)\,.
    \end{split}
\end{equation}
The equations of motions (EoMs) are given as
\begin{equation}\label{app:geom}
\begin{split}
&{{G}^{\mu}}_{\nu}+\alpha_2{{G_{(2)}}^{\mu}}_{\nu}={{T}^{\mu}}_{\nu}\,,\\
&\nabla^{\mu}\psi \nabla_{\mu}\psi-\frac{dV(\psi)}{d\psi} =0\,.
\end{split}
\end{equation}

\subsection{Comparison with Cosmological Billiards}
For general ``$d+1$" decompose form with shift vector field $N^i=0$, the metric reads
\begin{equation}
    ds^2=-N(t)^2dt^2+h_{ij}(t)dx^idx^j\,.
\end{equation} 
Based on \cite{Ivashchuk:2009hi}, we can get the effective action of the Einstein-GB-Scalar theory~\eqref{app:GB_lag}:
\begin{equation}
    \begin{aligned}
        S&=\int dt\tilde{N}^{-1}\left[\frac{1}{4}{A^i}_j{A^j}_i-\frac{1}{4}A^2+\frac{1}{2}{(\frac{d\psi}{dt})}^{2}\right]-\tilde{N}|h|V(\psi)\\
        &+\alpha_2|h|^{-1}\tilde{N}^{-3}\left[\frac{1}{8}A^2{A^i}_j{A^j}_i-\frac{1}{48}A^4-\frac{1}{6}A{A^i}_j{A^j}_k{A^k}_i\right.\\
        &\left.-\frac{1}{16}({A^i}_j{A^j}_i)^2+\frac{1}{8}{A^i}_j{A^j}_k{A^k}_l{A^l}_i\right]\,,
\end{aligned}
\end{equation}
where $|h|$ is the determinant of the spatial metric $h_{ij}$, $\tilde{N}=\frac{N}{\sqrt{|h|}}$, ${A^i}_j=h^{ik}\frac{d h_{kj}}{dt}$, and $A={A^i}_i$. Due to higher-curvature term, there are higher-order nonlinear velocity-dependent terms. Specifically, substituting the diagonal form
\begin{equation}
    ds^2=-e^{-2\beta(t)}dt^2+\sum_{i=1}^d e^{-2\beta^i(t)}dx_i^2,\quad \beta=\sum_{i=1}^{d}\beta^i\,,
\end{equation}
we obtain the effective Lagrangian (4) in the main text.

Moreover, the ``zero-mass constraint"\cite{Damour:2002et} is given by
\begin{equation*}
\begin{aligned}
    &\frac{1}{4}{A^i}_j{A^j}_i-\frac{1}{4}A^2+\frac{1}{2}{(\frac{d\psi}{dt})}^2+|h|\tilde{N}^2V(\psi)\\
    &+\frac{\alpha_2}{\tilde{N}^2|h|}
    \left[\frac{3}{8}A^{2}{A^i}_j{A^j}_i-\frac{1}{16}A^4-\frac{1}{2}A{A^i}_j{A^j}_k{A^k}_i\right.\\
    &\left.-\frac{3}{16}({A^i}_j{A^j}_i)^2+\frac{3}{8}{A^i}_j{A^j}_k{A^k}_l{A^l}_i\right]=0\,.
\end{aligned}
\end{equation*}
Under the given constraint, the field evolution ceases to follow the standard picture of free-particle motion in the curved target space of 
Cosmological Billiards~\cite{Damour:2002et}.

\subsection{Hairy black holes in Einstein-GB theory}\label{SUPPLEMENTARY_MATERIAL_GB}
We begin with the Kasner-like geometry given in (6) of the main text. The GB contribution to the EoMs reads
\begin{equation}\label{G(2)}
\begin{split}
{{G_{(2)}}^{t}}_{t}=&-(d-1)p_x(d p_x-4)\frac{(d-2)(d-3)p_x^2}{2 \tau^{4}}\,,\\
{{G_{(2)}}^{\tau}}_{\tau}=&-(d-1)p_x(4p_t+(d-4) p_x)\frac{(d-2)(d-3)p_x^2}{2 \tau^{4}}\,,\\
{{G_{(2)}}^{x}}_{x}=&-[4p_t(p_t+(d-2)p_x-3)\\&+(d-4)p_x((d-1)p_x-4)]\frac{(d-2)(d-3)p_x^2}{2 \tau^{4}}\,,
\end{split}
\end{equation}
where the subscript (2) denotes the second-order curvature invariant. The Einstein tensor from the Einstein–Hilbert part becomes
\begin{equation}\label{G}
\begin{split}
{{G}^{t}}_{t}=&-\frac{1}{2\tau^2}(d-1)p_x(dp_x-2)\,,\\
{{G}^{\tau}}_{\tau}=&-\frac{1}{2\tau^2}(d-1)p_x(2p_t+(d-2)p_x)\,,\\
{{G}^{x}}_{x}=&-\frac{1}{2\tau^2}[2p_t(p_t+(d-2)p_x-1)\\&+(d-2)p_x((d-1)p_x-2))]\,.
\end{split}
\end{equation}
Our subsequent analysis of Kasner eons is guided by the specific constraints of the system under consideration.

For the purpose of numerical study, we adopt plane‑symmetric AdS black holes as a prototypical system.
\begin{equation}\label{EQ:Ansatz}
   \mathrm{d}s^2=\frac{1}{z^2}\left[-f(z)\mathrm{e}^{-\chi(z)}\mathrm{d}t^2+\frac{\mathrm{d}z^2}{f(z)}+\mathrm{d}\vec{x}_{d-1}^2\right],\; \psi=\psi(z)\,.
\end{equation}
We set the boundary at $z=0$ and the singularity at $z\rightarrow\infty$, though their precise locations are immaterial to our analysis. Moreover, the horizon $z_H$ is determined by $f(z_H)=0$. By plugging the ansatz~\eqref{EQ:Ansatz} into~\eqref{app:geom}, we obtain the following three independent equations:
\begin{equation}\label{EQ:FulEoMz}
\begin{split}
&\ (\frac{1}{f}-2(d-2)(d-3)\alpha_2)(\frac{f'}{f}-\frac{d}{2z}-\frac{\chi'}{2})\\
&-\frac{d}{2zf}-\frac{V(\psi)}{(d-1)zf^2}=0\,,\\
&\ (d-1)(\frac{1}{f}-2(d-2)(d-3)\alpha_2)\chi'-\frac{z{\psi'}^2}{f}=0\,,\\
&\ \psi''+(\frac{z f'}{f}-\frac{z \chi'}{2}-(d-1))\frac{\psi'}{z}-\frac{1}{z^2f}\frac{dV}{d\psi}=0\,,
\end{split}
\end{equation}
where the prime denotes $z$-derivative. 

The coupled equations generally lack analytical solutions and must be solved numerically. Regularity at the horizon requires all functions to be finite and admit a Taylor expansion in $(z-z_H)$. Substituting this expansion into~\eqref{EQ:FulEoMz} reveals three independent horizon parameters: $z_H,\psi(z_H)$ and $\chi(z_H)$. Moreover, note that the EoMs~\eqref{EQ:FulEoMz} have two scaling symmetries:
\begin{equation}\label{scaling symmetry}
\begin{split}
  &z\rightarrow\lambda z, \quad (f, \chi,\psi)\rightarrow (f,\chi,\psi)\,,\\
&\chi\rightarrow \chi+\lambda,\quad (z, f,\psi)\rightarrow (z, f,\psi)\,, 
\end{split}
\end{equation}
with $\lambda$ constant. Taking advantage of the two scaling symmetries, we set $z_H=1$ and $\chi(z_H)=0$ for performing numerics. With these choices, the EoMs~\eqref{EQ:FulEoMz} are numerically determined once a value for $\psi(z_H)$. All black hole solutions presented in this work are obtained using this approach. The same method applies to other higher‑derivative corrected gravity theories—such as Lovelock gravity discussed later—provided that the EoMs remain second‑order.

Before concluding this section, we also express the equation in terms of $\rho = \ln z$. This coordinate transformation makes the periodic behavior of the relevant functions more evident for specific forms of the potential $V(\psi)$.
\begin{equation}\label{EQ:FulEoMrho}
\begin{split}
&\ (\frac{1}{f}-2(d-2)(d-3)\alpha_2 )(\frac{\dot{f}}{f}-d-\frac{\dot{\chi}}{2})\\&-d(d-2)(d-3)\alpha_2 -\frac{V(\psi)}{(d-1)f^2}=0\,,\\
&\ (d-1)(\frac{1}{f}-2(d-2)(d-3)\alpha_2 )\dot{\chi}-\frac{{\dot{\psi}}^2}{f}=0\,,\\
&\ \ddot{\psi}+(\frac{\dot{f}}{f}-d-\frac{\dot{\chi}}{2})\dot{\psi}-\frac{1}{f}\frac{dV}{d\psi}=0\,,
\end{split}
\end{equation}
where the dot denotes the derivative with respect to $\rho$. 
\begin{figure}[hpt]
    \centering
    \includegraphics[width=0.48\textwidth]{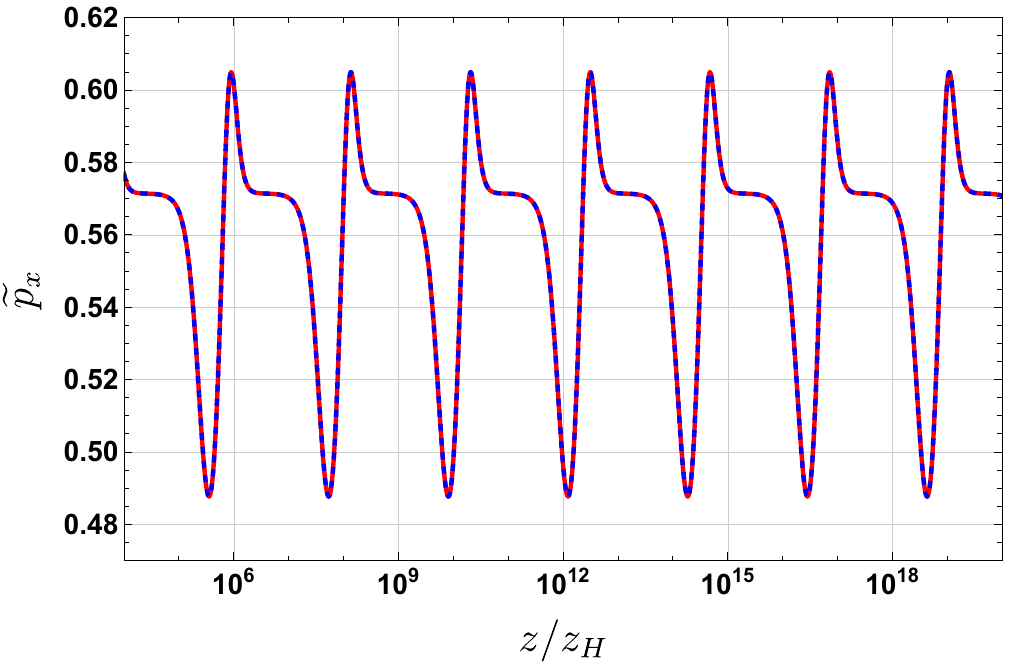}
    \caption{Far-interior evolution of $\widetilde{p}_x$ in the range $z/z_H\in[10^4,10^{20}]$ obtained from the full EoMs~\eqref{EQ:FulEoMrho} (solid red curve) and from the approximate equations (11) in the main text (dashed blue curve). The results are in quantitative agreement. We consider $V(\psi)=-42-3\psi^2+2\psi^4$ with $d=7$, $\alpha_2=10^{-2}$, and $\psi(z_H)=0.10$.}
    \label{app:comparion} 
\end{figure}

For the critical scenario with a quartic potential 
$V\sim c_4\psi^4$, our numerical simulations confirm that
$f\gg1$ and $\psi\gg1$ deep in the interior. This justifies neglecting the $1/f$ term in the first two equations of~\eqref{EQ:FulEoMrho}, leading to the simplified system given in (11) of the main text. The validity of this approximation is supported by Fig.~\ref{app:comparion}, which shows quantitative agreement between the dynamics derived from the full EoMs~\eqref{EQ:FulEoMrho} and those obtained from the approximate equations (11).

%
%
\subsection{Dynamics driven by scalar potential}\label{beyond_kasner}
In this section, we examine in detail the oscillatory behavior that arises when the potential $V(\psi)$ is not subdominant. These oscillations exhibit a rich and complex structure, being highly sensitive to initial conditions and strongly model-dependent. Nonetheless, through extensive numerical simulations, we have identified several common features that persist across different scenarios.
\begin{figure}[hpt]
    \centering
    \includegraphics[width=0.48\textwidth]{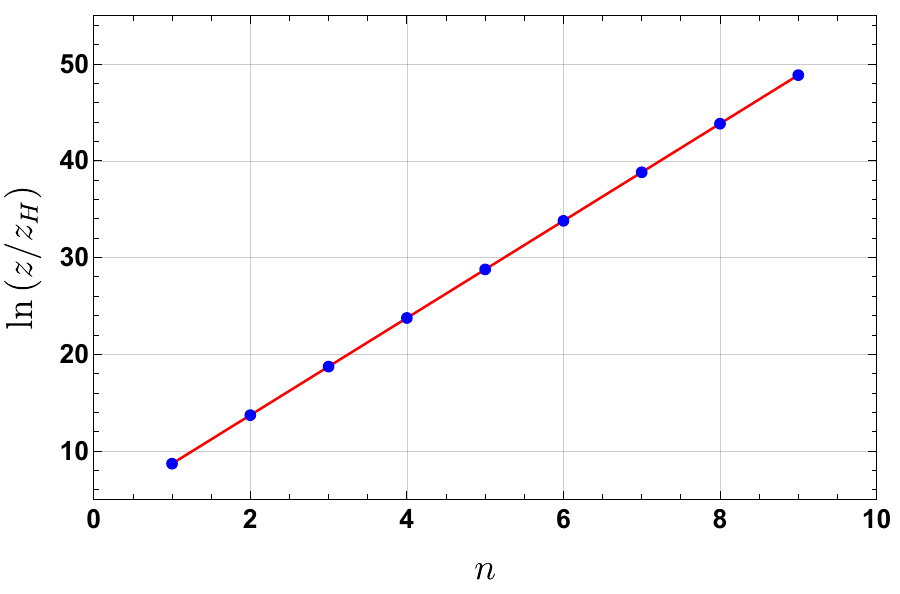}
    \caption{Positions of the $n$-th interior peak in a hairy Einstein–Gauss‑Bonnet black hole with $V(\psi)=-42-3\psi^2+2\psi ^4$, $d=7$, $\alpha_2=10^{-2}$, and $\psi(z_H)=0.10$. Blue points mark the positions extracted from the bottom panel of Fig.~\ref{fig:case12}; the red line shows a linear fit in the index $n$. The fit $\ln{(z/z_H)}=3.67+5.02n$ confirms a logarithmic periodicity.}
    \label{fig:caseII_period} 
\end{figure}

Our numerical analysis begins with the oscillatory regime of Case II of the critical scenario $(V\sim\psi^4)$. The bottom panel of Fig.\,1 (main text) implies a logarithmic scaling of the period. We test this by extracting the oscillation peaks, specifically the nine central peaks in the range $z/z_H\in[10^3,10^{22}]$ (blue points, Fig.~\ref{fig:caseII_period}). The numerical results give $\ln{(z_n/z_H)}=3.67+5.02n$ for the $n$-peak position $z_n$, from which we derive a logarithmic period given by $\ln{(z_{n+1}/z_H)}-\ln{(z_n/z_H)}=5.02$.
\begin{figure*}[hpt]
    \centering
        \includegraphics[width=0.46\textwidth]{fig/F3_n=6_caseII_p_liyx_260121.pdf} 
        \includegraphics[width=0.52\textwidth]{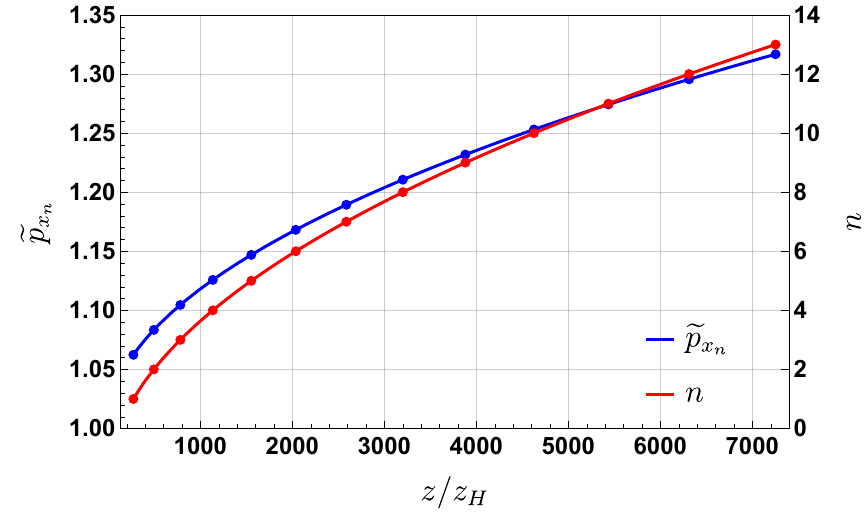}  \\
        
        \includegraphics[width=0.46\textwidth]{fig/F3_cosh_caseII_p_liyx_260122.pdf} 
        \includegraphics[width=0.52\textwidth]{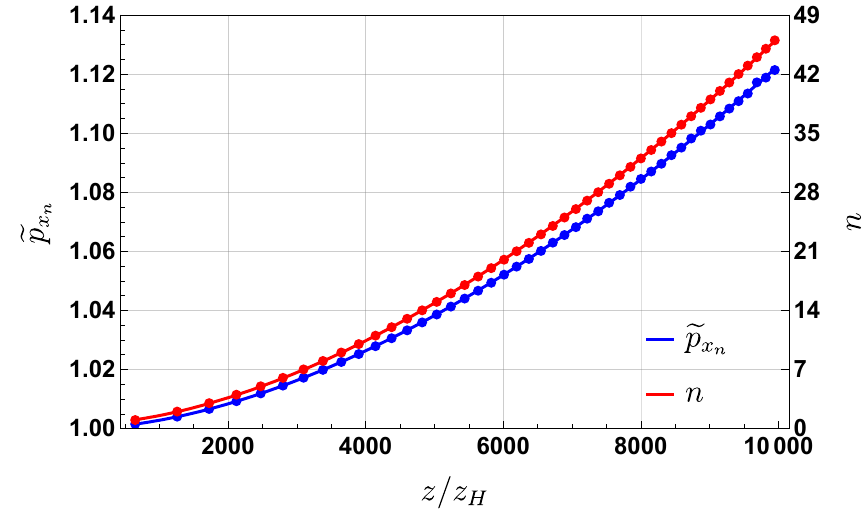}\\
      
        \includegraphics[width=0.46\textwidth]{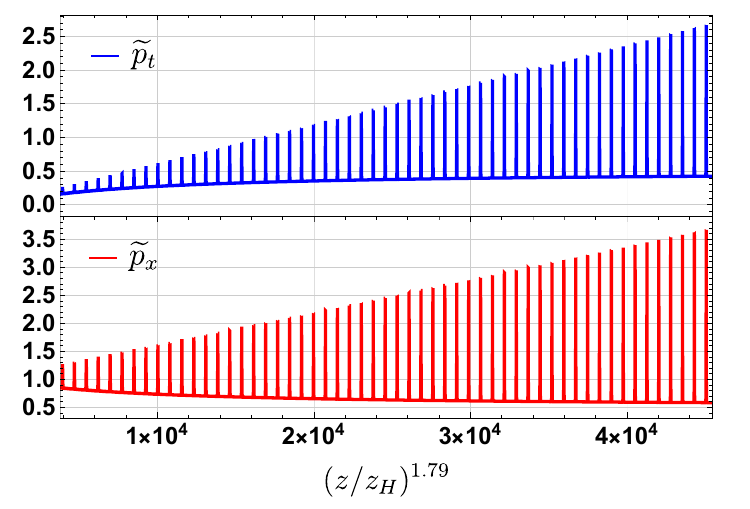} 
        \includegraphics[width=0.52\textwidth]{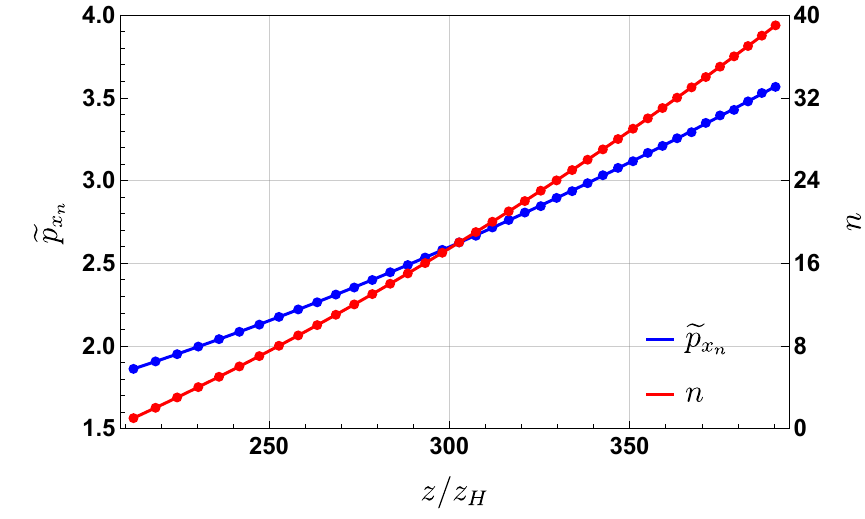}
    \caption{Interior spike dynamics driven by scalar potentials in Einstein–GB gravity.
    \textbf{Left panels}: Dynamics of the effective Kasner exponents $\widetilde{p}_t$ and $\widetilde{p}_x$ for $d=4$ and $\alpha_2=10^{-2}$ under three models:
    (i) Potential $V(\psi)=-12-\frac{3}{2}\psi^2+2\psi^6$ with $\psi(z_H)=0.30$;
    (ii) Potential $V(\psi)=-12-\frac{3}{2}\psi^2+\frac{1}{10}\cosh(\psi)$ with $\psi(z_H)=0.001$, and
    (iii) Potential $V(\psi)=-12-\frac{3}{2}\psi^2+\frac{1}{10}\cosh(\psi^2)$ with $\psi(z_H)=0.10$.
    \textbf{Right panels}: Fits of the $n$-th spike positions $z_n$ and amplitude $\widetilde{p}_{x_n}$ to their corresponding index $n$ (taken from the corresponding left panels): 
    (i)  $z_n=32.67{(n-0.09)}^{1/0.50}$, $\widetilde{p}_{x_n}=1.00+3.72*10^{-3}z_n^{0.50}$;  
    (ii) $z_n=1025.59{(n-0.56)}^{1/1.68}$, $\widetilde{p}_{x_n}=1.00+2.10*10^{-8}z_n^{1.69}$; 
    (iii) $z_n=40.77{(n+6.17)}^{1/{1.79}}$, $\widetilde{p}_{x_n}=1.01+5.18*10^{-5}z_n^{1.81}$.}
    \label{fig:spike_model}
\end{figure*}

For potential functions that diverge faster than $\psi^4$, the system exhibits sustained oscillatory growth. This spike dynamics introduces considerable complexity. Despite the increased mathematical intricacy of the EoMs, numerical solutions still allow us to extract insights into the period and amplitude. We follow an approach analogous to that in critical Case II. First, we extract the positions $z_n$ of successive peaks. If these follow the relation
\begin{equation}\label{ngamma}
z_n={a_1(n+b_1)}^{1/{\gamma}}\,,
\end{equation}
for constants $a_1$, $b_1$ and $\gamma$, then the spacing between successive peaks obeys
\begin{equation}
z_{n+1}^{\gamma}-z_{n}^{\gamma}=a_1\,,
\end{equation}
indicating a power-law periodicity. Second, the amplitudes can be analyzed similarly. Using $\widetilde{p}_{x}$ as an example, if the amplitude $\widetilde{p}_{x_n}$ of the $n$-th peak fits the form
\begin{equation}\label{pdelta}
\widetilde{p}_{x_n}=a_2z_n^{\delta}+b_2\,,
\end{equation}
for constants $a_2$, $b_2$ and $\delta$, then the peak amplitudes exhibit a corresponding power-law scaling.

\begin{figure*}[htbp]
    \centering
        \includegraphics[width=0.46\textwidth]{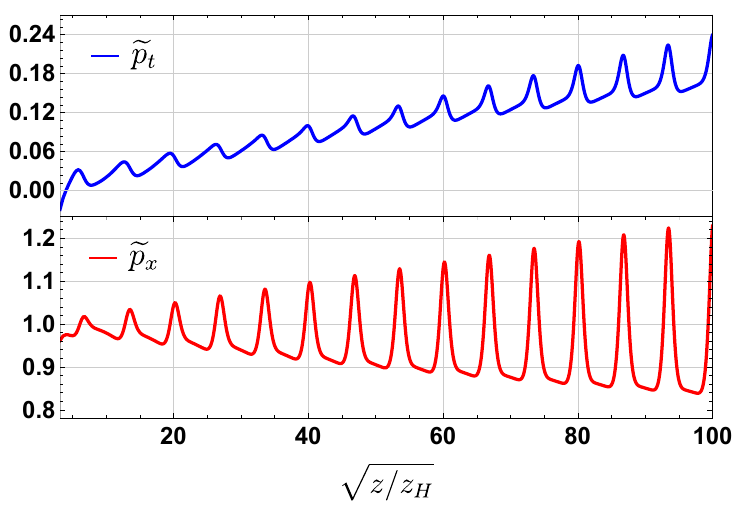} 
        \includegraphics[width=0.52\textwidth]{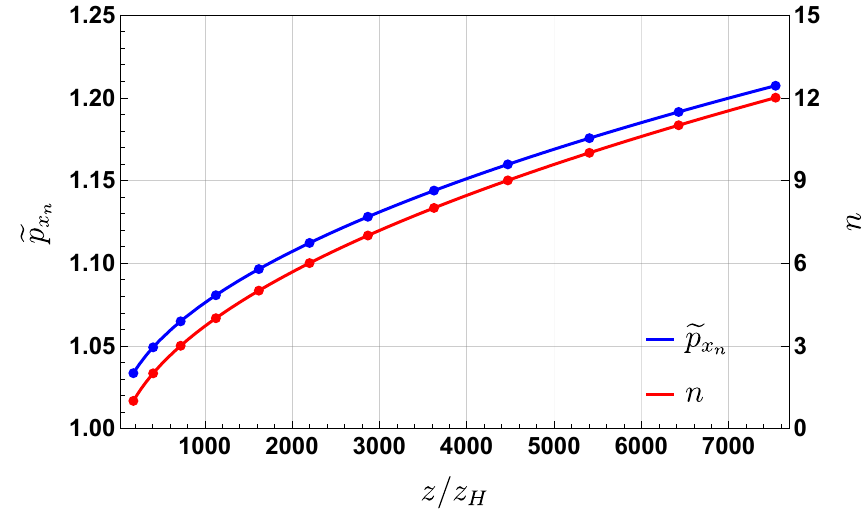}  \\
        
        \includegraphics[width=0.46\textwidth]{fig/F3_n=6_caseII_p_liyx_260121.pdf} 
        \includegraphics[width=0.52\textwidth]{fig/F5_spike_gb_phi6_0.3.pdf}
    \caption{Interior spike dynamics under different initial conditions for $V(\psi)=-12-\frac{3}{2}\psi^2+2\psi^6$ in Einstein–GB gravity. We choose $d=4$ and $\alpha_2=10^{-2}$.
    \textbf{Left panels}: Evolution of the effective Kasner exponents $\widetilde{p}_t$ and $\widetilde{p}_x$ for two initial horizon values of the scalar field: 
    (i) $\psi(z_H)=0.10$;
    (ii) $\psi(z_H)=0.30$.
    \textbf{Right panels}: Fits of the $n$-th spike positions $z_n$ and amplitude $\widetilde{p}_{x_n}$ to their corresponding index $n$ (taken from the corresponding left panels): 
    (i) $z_n=32.67{(n-0.09)}^{1/0.50}$, $\widetilde{p}_{x_n}=0.99+2.64*10^{-3}z_n^{0.48}$;
    (ii) $z_n=44.68{(n+0.04)}^{1/0.50}$, $\widetilde{p}_{x_n}=1.00+3.72*10^{-3}z_n^{0.50}$. }
    \label{fig:spike_initial}
\end{figure*}

The scaling relations identified above are supported by our numerical simulations. We also find that the emergent spike dynamics are critically dependent on the specific model potential. This is illustrated in the left panels of Fig.~\ref{fig:spike_model} for potentials with power-law, exponential, and super-exponential divergence. In all cases, the oscillations exhibit a power-law periodicity in $z/z_H$, described by~\eqref{ngamma}, although the exponent $\gamma$ varies with $V(\psi)$. Consequently, the late-time behavior of $\widetilde{p}_x$ is periodic in $(z/z_H)^\gamma$ with a power-law modulated amplitude, as captured by the fit to~\eqref{pdelta}. Moreover, the detailed spike dynamics exhibit sensitivity to the initial conditions. As shown in Fig.~\ref{fig:spike_initial} for $V(\psi)=-12-\frac{3}{2}\psi^2+2\psi^6$, although the oscillations in both cases retain a periodicity with square-root scaling, the precise locations and amplitudes of the peaks vary with different values of $\psi(z_H)$. An analytical understanding of these rich behaviors remains an interesting open question, which we defer to future work.

\section{Higher-order Lovelock gravity}\label{SM-LL}
Many higher-curvature gravity theories suffer from potential pathologies, typically leading to EoMs of fourth order or higher. In contrast, Lovelock gravity stands out as the most general theory whose field equations remain second order, making it a natural and well-behaved extension of Einstein gravity to higher dimensions. This property establishes Lovelock gravity as a valuable laboratory for studying the genuine effects of higher-curvature terms on gravitational dynamics. In $(d+1)$-spacetime dimensions, the Lovelock terms read~\cite{Lovelock:1971yv,Maeda:2011ii}:
\begin{equation*}
\mathcal{L}_{Lovelock}=\sum_{N=2}^{[(d+1)/2]}\alpha_N\mathcal{L}_{(N)}\,,\\
\end{equation*}
\begin{equation*}
\mathcal{L}_{(N)}=\frac{1}{2^N}\delta^{\alpha_1\beta_1...\alpha_N\beta_N}_{\mu_1\nu_1...\mu_N\nu_N}{R^{\mu_1\nu_1}}_{\alpha_1\beta_1}...{R^{\mu_N\nu_N}}_{\alpha_N\beta_N}\,,
\end{equation*}
%
%
where the $\delta$ symbol denotes a totally anti-symmetric product of Kronecker deltas. In even dimensions, the contribution to the action of the $(d+1)/2$-th order Lagrangian constitutes a topological invariant and consequently does not contribute to the field equations. For example, the GB case corresponds to $N=2$ and becomes a topological invariant in $(3+1)$ dimensions.

As in the main-text GB example, a critical dynamics appears once the scalar potential scales as $\mathcal{O}(V)\sim\mathcal{O}(\psi^{\frac{2N}{N-1}})$. This modification of the Kasner eons leads to persistent oscillatory behavior. 
To demonstrate the generality of this phenomenon, we consider third-order Lovelock gravity as an example. Since the contributions from Einstein gravity and the GB term have been treated in Section~\ref{SUPPLEMENTARY_MATERIAL_GB}, we now focus solely on computing the effects arising from the cubic Lovelock terms. The cubic curvature terms read
\begin{equation}
    \begin{split}
       \mathcal{L}_{3}=&R^3+24R^{\mu\nu\alpha\beta}R_{\alpha\mu}R_{\beta\nu}+24R^{\mu\nu\alpha\beta}R_{\alpha\beta\nu\rho}{R^{\rho}}_{\mu}\\
        &+8{R^{\mu\nu}}_{\alpha\rho}{R^{\alpha\beta}}_{\nu\rho}{R^{\rho\sigma}}_{\mu\beta}+2R_{\alpha\beta\rho\sigma}R^{\mu\nu\alpha\beta}{R^{\rho\sigma}}_{\mu\nu}\\
        &+3RR^{\mu\nu\alpha\beta}R_{\mu\nu\alpha\beta}-12RR^{\mu\nu}R_{\mu\nu}+16R^{\mu\nu}R_{\nu\alpha}{R^{\alpha}}_\mu\,.
    \end{split}
\end{equation}
As argued in the main text, for the Kasner eon geometry, they diverge much faster than the GB case as the singularity is approached. Thus, the cubic terms will dominate the interior dynamics at late interior times. 

Substituting into the Kasner econ geometry (6) of the main text, one has
\begin{equation}\label{LL_G(3)}
\begin{split}
{{G_{(3)}}^{t}}_{t}=&-\frac{(d-5)(d-4)(d-3)(d-2)p_x^4}{2\tau^6}\\
&\times(d-1)p_x(dp_x-6)\,,\\
{{G_{(3)}}^{\tau}}_{\tau}=&-\frac{(d-5)(d-4)(d-3)(d-2)p_x^4}{2\tau^6}\\
&\times(d-1)p_x(6p_t+(d-6)p_x)\,,\\
{{G_{(3)}}^{x}}_{x}=&-\frac{(d-5)(d-4)(d-3)(d-2)p_x^4}{2\tau^6}\,,\\
&\times[6p_t(p_t+(d-2)p_x-5)\\
&+(d-6)p_x((d-1)p_x-6)]\,.
\end{split}
\end{equation}
Diverging as $1/\tau^6$ when $\tau\rightarrow 0$, they are significantly faster than the GB case described by~\eqref{G(2)}.

The critical case now is obtained when the potential is cubic in $\psi$, for which its contribution becomes comparable to the kinetic term. For the Kasner eon in~(6) of the main text, the highest-order equation ${{G_{(3)}}^t}_t = {{G_{(3)}}^x}_x$ yields two families of Kasner exponents:
\begin{itemize}
    \item Case I: $p_t=p_x$, corresponding to a modified Kasner epoch in which the subdominant conditions ${{G_{(2)}}^t}_t = {{G_{(2)}}^x}_x$ and ${{G}^t}_t={{G}^x}_x$ are also satisfied. It corresponds to a stable hairy Kasner eon.
    
    \item Case II: $p_t=5-(d-1)p_x$, where the Kasner eon geometry becomes unstable and gives way to periodic oscillations. Specifically, in the presence of the GB, the anisotropy ${{G_{(2)}}^t}_t-{{G_{(2)}}^x}_x$ diverges and drives the Kasner eon breakdown. Notably, even in the absence of the GB term (pure Einstein gravity), the divergent anisotropy ${{G}^t}_t\neq {{G}^x}_x$ itself is sufficient to produce the same oscillatory regime.
\end{itemize}
\begin{figure}[h]
    \centering
    \includegraphics[width=0.48\textwidth]{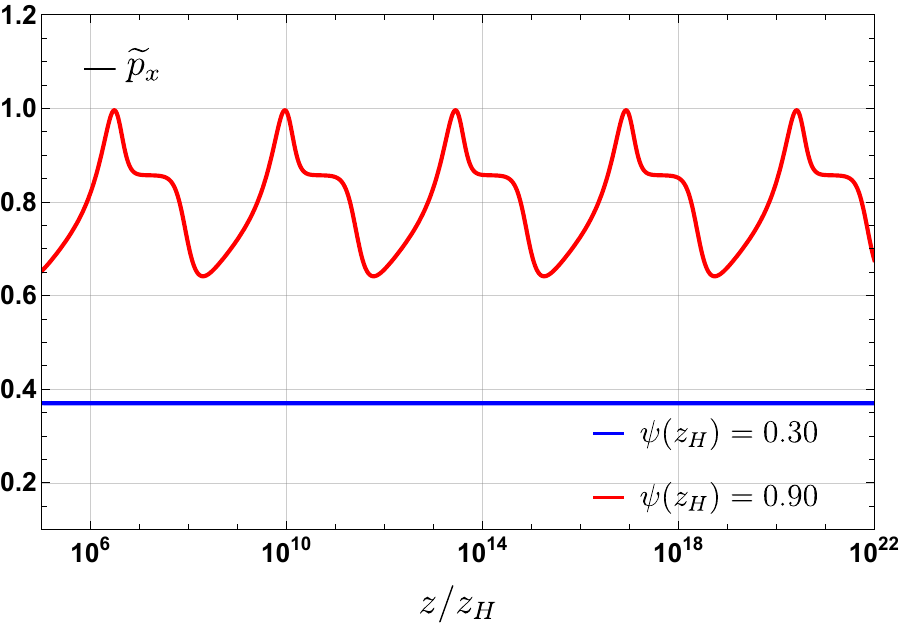}
    \caption{Far‑interior evolution of $\widetilde{p}_x$ obtained by solving the full EoMs in third-order Lovelock theory with $d=7$, $\alpha_2=10^{-2}$, $\alpha_3=10^{-2}$ and $V(\psi)=-42-\frac{3}{2}\psi^2+\psi^3$. The blue curve with $\psi(z_H)=0.30$ is for the modified Kasner eons (Case I) and the red curve $\psi(z_H)=0.90$ corresponds to periodic oscillations (Case II).  }
    \label{fig:R3} 
\end{figure}
The above behaviors are presented in Fig.~\ref{fig:R3} by solving the full EoMs:
\begin{equation}
\begin{split}
    &-(d-2)(d-3)(d-4)(d-5)\alpha_3((\frac{d}{z}+\frac{3\chi'}{2})f-3f')\\
    &+(\frac{1}{f}-2(d-2)(d-3)\alpha_2)(\frac{f'}{f}-\frac{\chi'}{2}-\frac{d}{2z})-\frac{d}{2zf}\\
    &=\frac{V(\psi)}{(d-1)zf^2}\,,\\
    &\ 3(d-1)(d-2)(d-3)(d-4)(d-5)\alpha_3f\chi'\\
    &+(d-1)(\frac{1}{f}-2(d-2)(d-3)\alpha_2)\chi'=\frac{z{\psi'}^2}{f}\,,\\
    &\ \psi''+(\frac{zf'}{f}-\frac{z\chi'}{2}-(d-1))\frac{\psi'}{z}-\frac{1}{z^2f}\frac{dV}{d\psi}=0\,,
\end{split}
\end{equation}
for the hairy black hole~\eqref{EQ:Ansatz} with the Einstein-GB-cubic Lovelock gravity coupled to a scalar field. Note that the above EoMs also enjoy the scaling symmetries~\eqref{scaling symmetry}. Moreover, after turning off $\alpha_3$, one recovers the EoMs for the Einstein-GB theory~\eqref{EQ:FulEoMz}.

In contrast to GB case, the boundary between the two cases is governed by higher-order EoM, which constrain the system parameters through higher-order matching conditions:
\begin{equation}\label{G(3)boundary}
\begin{split}
&\text{Case I}:\ c_3^2\alpha_3=\frac{(dp_x-3)^2}{216(d-5)(d-4)(d-3)(d-2)(d-1)p_x^5}\,,\\
&\text{Case II}:\ c_3^2\alpha_3=\frac{-(dp_x-6)^{-1}}{54(d-5)(d-4)(d-3)(d-2)(d-1)p_x^5}\,,
\end{split}
\end{equation}
where $\alpha_3$ is the coupling constant of third-order Lovelock terms and $c_3$ the coefficients of cubic potential $V\sim c_3 \psi^3$.
Unlike the phase diagram shown in Fig.\;2 of the main text, the phase diagram on the $\alpha_3$-$c_3$ plane exhibits an overlapping region between Case I (modified Kasner eon) and Case II (periodic oscillation). In the overlapping region, whether the system is in a Kasner eon or exhibits periodic oscillations depends on the choice of initial conditions (see also Fig.~\ref{fig:R3}).

\begin{figure*}[htbp]
    \centering
        \includegraphics[width=0.47\textwidth]{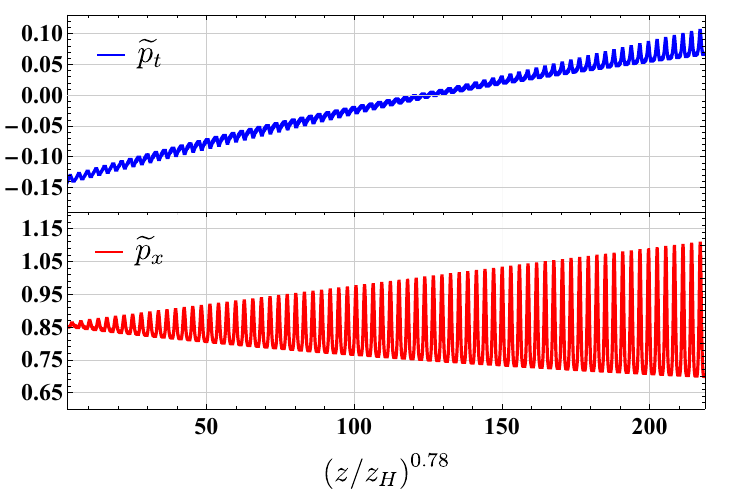} 
        \includegraphics[width=0.52\textwidth]{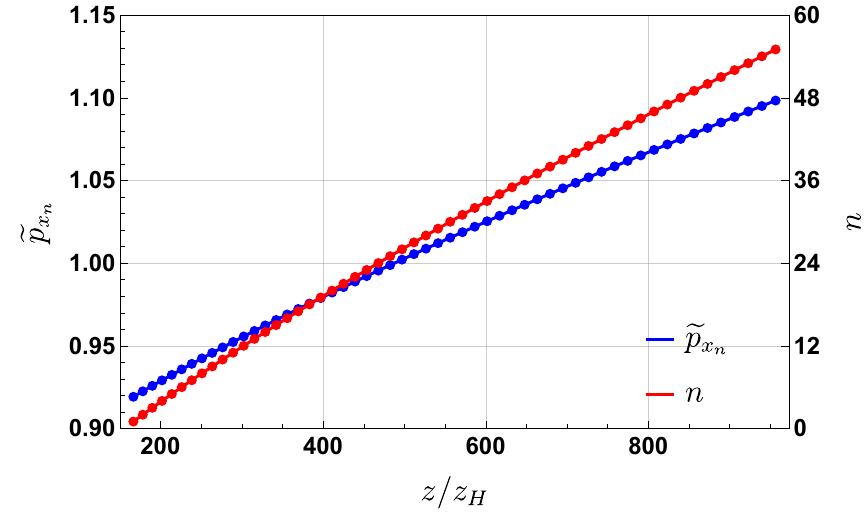}  \\
        
        \includegraphics[width=0.47\textwidth]{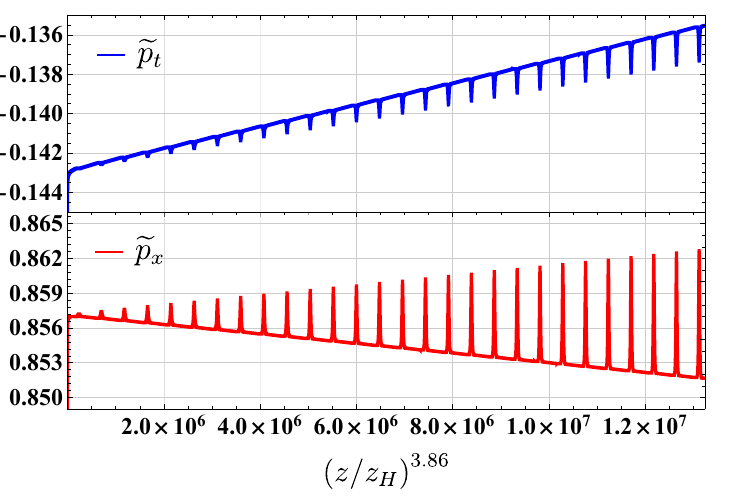} 
        \includegraphics[width=0.52\textwidth]{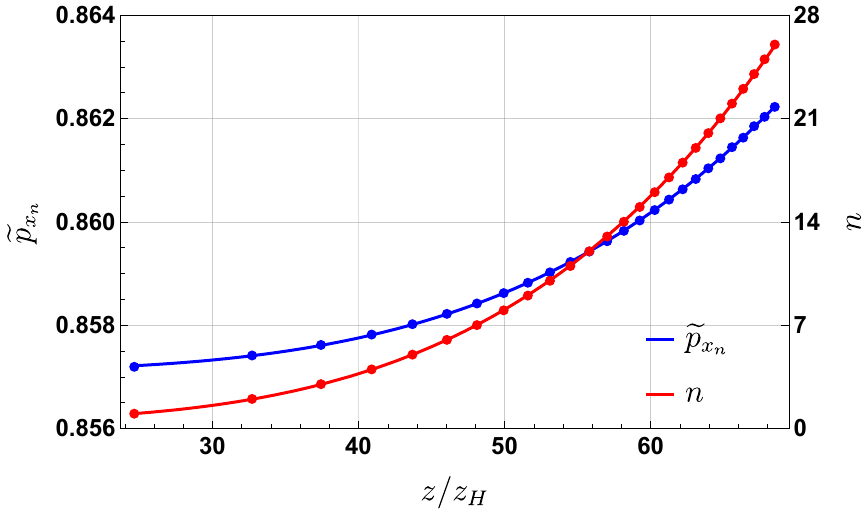}
    \caption{Spike dynamics across different scalar potentials in third-order Lovelock gravity with $d=7$. 
    \textbf{Left panels}: Dynamics of the effective Kasner exponents $\widetilde{p}_t$ and $\widetilde{p}_x$ for $d=7$ for 
    (i) Potential $V(\psi)=-42-\frac{3}{2}\psi^2+\psi^4$ with $\alpha_2=10^{-2}$, $\alpha_3=10^{-2}$, $\psi(z_H)=0.90$;
    (ii) Potential $V(\psi)=-42+\frac{1}{10}\cosh(\psi)$ with $\alpha_2=10^{-3}$, $\alpha_3=10^{-4}$, and $\psi(z_H)=0.01$.
    \textbf{Right panels}: Fits of the $n$-th spike positions $z_n$ and amplitude $\widetilde{p}_{x_n}$ to their corresponding index $n$ (taken from the corresponding left panels): 
    (i) $z_n=3.89{(n+0.64)}^{1/0.78}$, $\widetilde{p}_{x_n}=0.86+1.15*10^{-3}z_n^{0.78}$. 
    (ii) $z_n=29.63{(n-0.51)}^{1/3.86}$, $\widetilde{p}_{x_n}=0.86+3.50*10^{-10}z_n^{3.90}$.}
    \label{fig:n=4_spike}
\end{figure*}

Within third-order Lovelock theory, spike dynamics are observed when the scalar potential is characterized by a divergence faster than $\psi^3$. Fig.~\ref{fig:n=4_spike} illustrates these dynamics for representative power-law and exponential potentials, respectively. In contrast to periodic oscillation of Fig.~\ref{fig:R3}, the effective exponents $\widetilde{p}_t$ and $\widetilde{p}_x$ now exhibit sustained oscillatory growth as the singularity is approached, revealing a possible discrete self-similarity. In all cases, the oscillations exhibit a power-law periodicity in $z/z_H$, described by~\eqref{ngamma}, although the exponent $\gamma$ varies with $V(\psi)$. This similarity confirms that the qualitative behavior aligns with the patterns seen in the GB case.

\section{Finite-volume singularity driven by potentials unbounded from below}\label{SM-FS}
Our analysis of scalar-driven spike dynamics is restricted to potentials bounded from below. This restriction is necessary because a potential lacking a lower bound is well known to introduce severe theoretical inconsistencies. Specifically, we find that such an unbounded potential inevitably culminates in a finite-volume singularity. While similar singularities have been reported in pure Einstein-GB gravity with negative coupling (\emph{e.g.}~\cite{Bueno:2024fzg}), a healthy tree-level UV completion (free of ghosts and tachyons) forces the GB coupling to be nonnegative~\cite{Cheung:2016wjt}.

\begin{figure*}[htbp]
    \centering
    
         \includegraphics[width=0.42\textwidth]{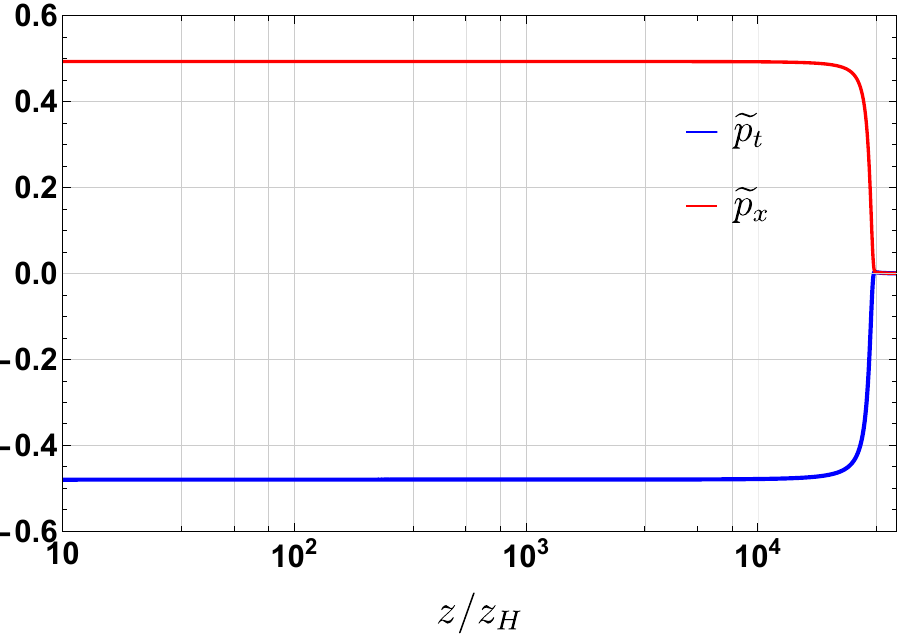} \quad
        \includegraphics[width=0.553\textwidth]{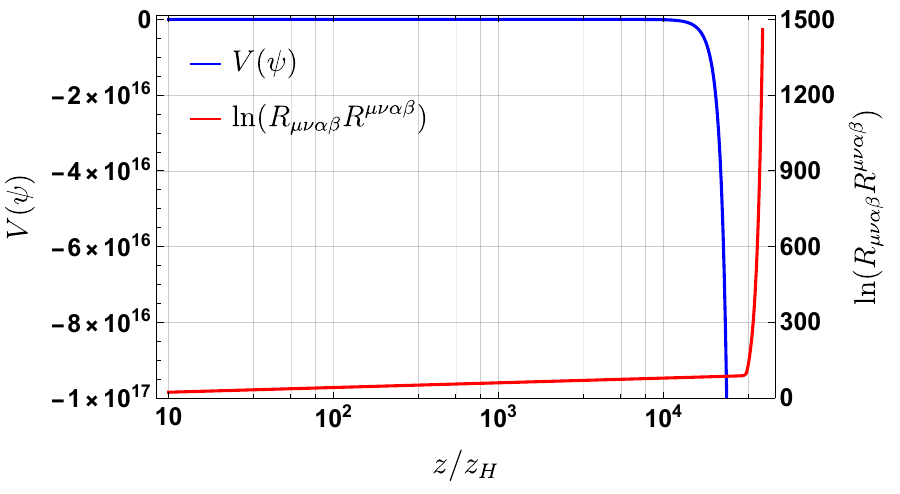}  \\
      
      \includegraphics[width=0.42\textwidth]{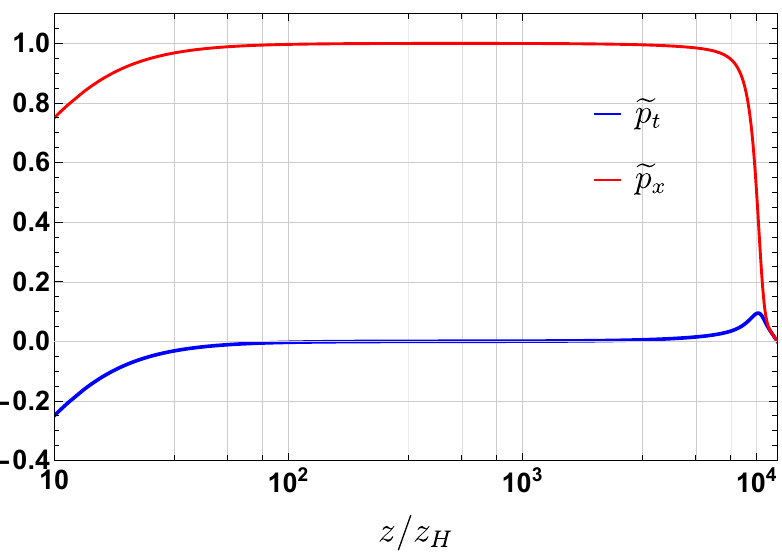} \quad
        \includegraphics[width=0.553\textwidth]{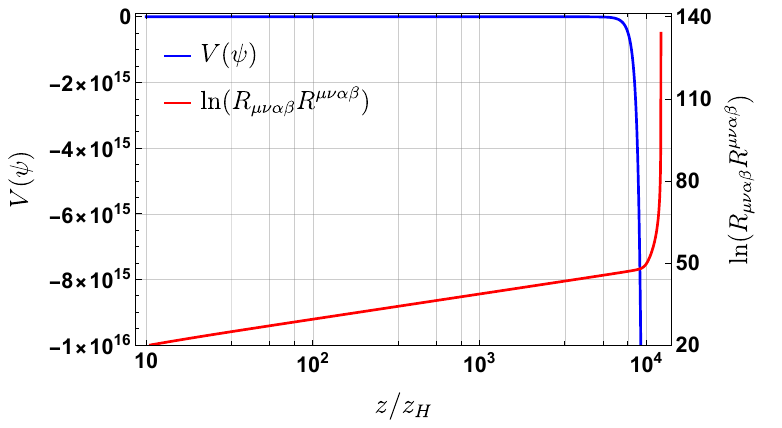}\\  
        
        \includegraphics[width=0.42\textwidth]{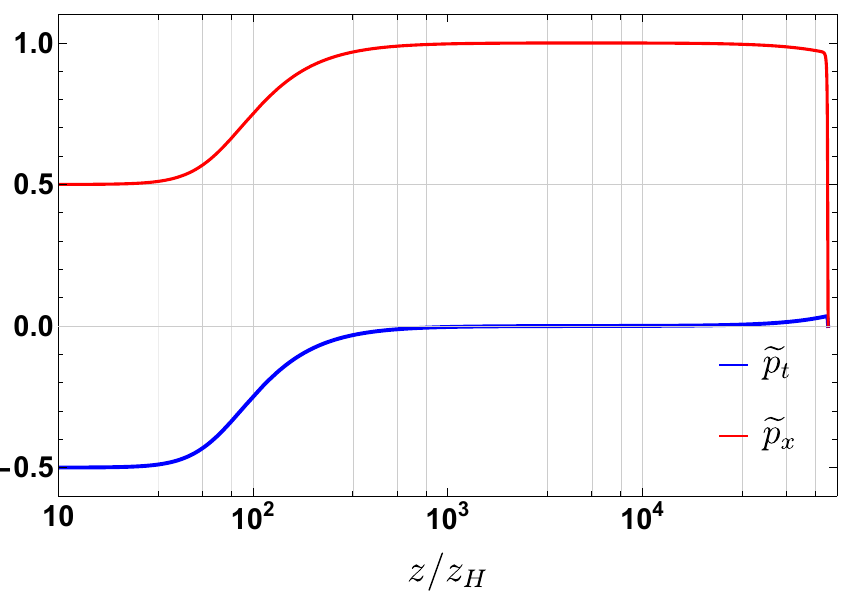} \quad
        \includegraphics[width=0.553\textwidth]{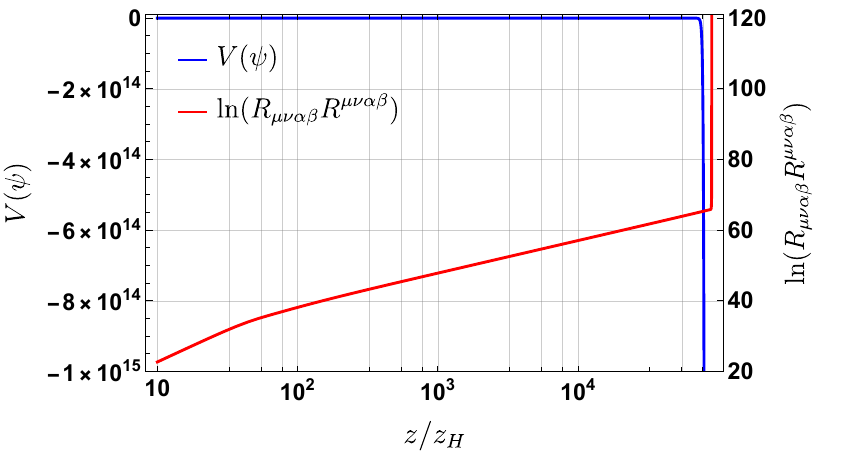} 
    \caption{Finite‑volume singularity driven by potentials unbounded from below. 
    \textbf{Left panels}: Dynamics of the effective Kasner exponents $\widetilde{p}_t$ and $\widetilde{p}_x$ for $d=4$ in three models: 
    (i) Einstein gravity, $V(\psi)=-11-\frac{3}{2}\psi^2-\cosh(\psi^2)$, with $\psi(z_H)=0.50$;
    (ii) Einstein-GB gravity, $V(\psi)=-12-\frac{3}{2}\psi^2-\frac{1}{10}\psi^6$, $\alpha_2=10^{-4}$, $\psi(z_H)=0.001$;
    (iii) Einstein-GB-cubic Lovelock gravity, $V(\psi)=-12-\frac{3}{2}\psi^2-\frac{1}{100} \cosh(\frac{\psi}{10})$, $\alpha_2=10^{-8}$, $\psi(z_H)=0.001$. 
    \textbf{Right panels}: The divergence of the potential $V(\psi)$ and of the Kretschmann scalar $R^{\mu\nu\alpha\beta}R_{\mu\nu\alpha\beta}$ in each model confirms that the potential is unbounded from below and that all such configurations develop a curvature singularity at finite volume.}
    \label{fig:catastrophe}
\end{figure*}

In Fig.~\ref{fig:catastrophe}, we present numerical results for three distinct gravitational models. From top to bottom, the panels correspond to: Einstein gravity with a super-exponential potential; Einstein-GB gravity with a power-law potential; and the Einstein-GB-cubic Lovelock gravity with an exponential potential. For each model, we show the evolution of $\widetilde{p}_x$, $\widetilde{p}_t$ and the Kretschmann scalar $R^{\mu\nu\alpha\beta}R_{\mu\nu\alpha\beta}$, together with $V(\psi)$. The left panels show the breakdown of the effective Kasner exponents across different gravitational models, indicating a violent departure from standard interior dynamics.
The right panels directly plot the divergence of the potential the Kretschmann scalar in each model. The simultaneous blow-up of both quantities confirms that a finite‑volume singularity readily occurs when the scalar potential is unbounded from below and sufficiently rapidly diverging. Crucially, this outcome is independent of the specific gravitational theory. In particular, this singularity already appears in Einstein gravity alone—demonstrating that it is intrinsic to the unbounded potential in the base theory, and not a consequence of the higher‑derivative corrections under investigation.

Numerical results demonstrate that in the vicinity of the singularity, both effective exponents $\widetilde{p}_x$, $\widetilde{p}_t$ rapidly approach zero—a behavior consistently observed across the different gravitational theories we have considered (see the left panels of Fig.~\ref{fig:catastrophe}). This raises the question of whether this represents a more universal phenomenon and whether it carries deeper physical significance, both of which merit further investigation.
%
\end{document}